\documentclass[11pt,a4paper,fleqn]{article}
\catcode`\@=11
\@addtoreset{equation}{section}
\def\theequation{\arabic{section}.\arabic{equation}}
\def\appendix{\renewcommand{\thesection}{\Alph{section}}\setcounter{section}{0}
             \renewcommand{\theequation}
           {\mbox{\Alph{section}.\arabic{equation}}}\setcounter{equation}{0}}

\def\maketitle{\thispagestyle{empty}\setcounter{page}0\newpage
               \renewcommand{\thefootnote}{\arabic{footnote}}
                 \setcounter{footnote}0}
\renewcommand{\thanks}[1]{\renewcommand{\thefootnote}{\fnsymbol{footnote}}
              \footnote{#1}\renewcommand{\thefootnote}{\arabic{footnote}}}

\renewcommand{\title}[1]{\begin{center}\Large\bf #1\end{center}\rm\par\bigskip}

\renewcommand{\author}[1]{\begin{center}\Large #1\end{center}}
\newcommand{\address}[1]{\begin{center}\large #1\end{center}}

\def\babs{\hrule\par\begin{description}\item{Abstract: }\it}
\def\eabs{\par\end{description}\hrule\par\medskip\rm}
\renewcommand{\date}[1]{\par\bigskip\par\sl\hfill #1\par\medskip\par\rm}

\newcommand{\dip}[1]{${}^{(#1)}$ Dipartimento di Fisica, Universit\`a di Trento\\
                                     via Sommarive 14, 38123 Trento, Italia\\}
\newcommand{\infn}[1]{${}^{(#1)}\,$TIFPA (INFN), Trento, Italia\\ \medskip}
\newcommand{\csic}[1]{${}^{(#1)}\,$Consejo Superior de Investigaciones Científicas\\
ICE (CSIC-IEEC), UAB Campus, 08193 Bellaterra, Barcelona, Spain \\}

\newcommand{\guido}[1]{Guido Cognola${}^{#1}$\thanks{e-mail:\sl cognola@science.unitn.it\rm}}
\newcommand{\sergio}[1]{Sergio Zerbini${}^{#1}$\thanks{e-mail:\sl zerbini@science.unitn.it\rm}}

\newcommand{\emilio}[1]{Emilio Elizalde${}^{#1}$\thanks{e-mail:\sl elizalde@ieec.uab.es\rm}}

\newcommand{\s}[1]{\section{#1}}
\renewcommand{\ss}[1]{\subsection{#1}}

\def\hs{\qquad}               
\def\nn{\nonumber}            
\def\beq{\begin{eqnarray}}    
\def\be{\begin{eqnarray}}
\def\eeq{\end{eqnarray}}      
\def\ee{\end{eqnarrayn}}
\def\at{\left(}               
\def\aq{\left[}               
\def\ag{\left\{}              
\def\cp{\right.}              
\def\ct{\right)}              
\def\cq{\right]}              

\def\R{{\hbox{{\rm I}\kern-.2em\hbox{\rm R}}}}   
\def\H{{\hbox{{\rm I}\kern-.2em\hbox{\rm H}}}}   
\def\N{{\hbox{{\rm I}\kern-.2em\hbox{\rm N}}}}   
\def\C{{\ \hbox{{\rm I}\kern-.6em\hbox{\bf C}}}} 
\def\Z{{\hbox{{\rm Z}\kern-.4em\hbox{\rm Z}}}}   
\def\ii{\infty}                                  
\def\Res{\mathop{\rm Res}\nolimits}                
\def\res{\mathop{\rm res}\nolimits}                
\renewcommand{\Re}{\mathop{\rm Re}\nolimits}       
\def\dir{/\kern-.7em D\,}                          
\def\lap{\Delta\,}                                 
\def\al{\alpha}
\def\ga{\gamma}

\def\ze{\zeta}

\def\la{\lambda}

\def\Ga{\Gamma}

\def\Om{\Omega}

\def\be{\begin{equation}}
\def\ee{\end{equation}}
\def\bea{\begin{eqnarray}}
\def\eea{\end{eqnarray}}

\def\nn{\nonumber}
\def\e{{\rm e}}

\renewcommand{\title}[1]{\begin{center}\Large\bf #1\end{center}\rm\par\bigskip}
\renewcommand{\author}[1]{\begin{center}\Large #1\end{center}}

\begin{document}

\title{Functional Determinant of the Massive Laplace Operator \\ and the Multiplicative Anomaly}  
\author{\guido{a,b}, \emilio{c}, and \sergio{a,b}}
\address{\dip{a}\infn{b}\csic{c}}

\begin{abstract}
After a brief survey of zeta function regularization issues and of the related multiplicative anomaly, illustrated with a couple of basic examples, namely the harmonic oscillator and quantum field theory at finite temperature, an application of these methods to the computation of functional determinants corresponding to massive Laplacians on  spheres in arbitrary dimensions is presented. Explicit formulas are provided for the Laplace operator on spheres in $N=1,2,3,4$ dimensions and for `vector' and `tensor' Laplacians on the unitary sphere $S^4$. 
\end{abstract}


\section{Introduction}
 
In quantum field theory (QFT), the Euclidean partition function plays a very important role. The full propagator and all other $n-$point correlation functions can be computed by means of it. Moreover, this tool can be extended without problem to curved space-time \cite{buch}.
As a formalism this is extremely beautiful but it must be noticed that in relativistic  quantum  field theories an infinite number of degrees of 
freedom is involved and, as a consequence, ultraviolet divergences will be present, thus rendering regularization and renormalization compulsory.    

In the one-loop approximation, and in the external field approximation too, one may
describe (scalar) quantum fields by means of a (Euclidean) path integral and
express the Euclidean partition function as a function of functional determinants associated with the differential operators involved. In this way,  the partition function reads
\beq
Z \simeq \left( \det L \right)^{-1/2}\,,
\eeq
with $L$ being an elliptic self-adjoint non-negative operator, the small fluctuation operator. We thus see that the computation of 
Euclidean one-loop partition functions reduces to the computation of functional determinants.
 
As already mentioned, functional determinants are divergent quantities which need to be regularized. For a long period this was performed in the physics literature case by case, by adding `reasonable' correcting terms to formulas such as 
$\det (AB) =(\det A)(\det B)$, which did look suspect (see, e.g. \cite{dea1}, among very many refs)---and are, in fact, generically {\it wrong}. No wonder, since those $\det$'s are in {\it no way} regular determinants, but {\it regularized}  ones, which do not satisfy the usual properties of det's, in particular, the multiplicative property. \footnote{The same happens with regularized traces which, to begin with, are non-linear, generically.} 

Soon after this was clearly understood, some seminal works appeared  \cite{wo87,ka89,kv94} (see also \cite{eli1}, and \cite{toms} and references therein) where the already existing rigorous, simple, and also very beautiful mathematical formulation of the so-called Wodzicki's or residue calculus for pseudo-differential operators was put forward and made explicit for use in theoretical physics. However, many practitioner physicists are still now unaware of these fundamental methods. In special, the so-called {\it multiplicative anomaly} or {\it defect} of the zeta-regularized determinant (a well-established definition stemming from Atiyah, Ray and Singer \cite{ars1}) is a perfectly-under-control quantity which can be given by a very simple formula in terms of the Wodzicki residue, which is on its turn the {\it only} true trace (up to a multiplicative constant) one can define on the whole class of pseudo-differential operators---and extends, in a unique way, the Dixmier trace and the Adler-Manin one, which are just particular cases of it. This is, in a word, standard theory since the early 90's, at the very least.\footnote{There is nothing mysterious, uncontrolled, or even difficult, in this matter, contrary to the impression one may get from the many papers around carrying out ad-hoc calculations for each particular situation.}  Concerning the spherical case in which we will be particularly interested in this paper, a series of notorious contributions in this respect (see \cite{hk1}, and \cite{dowker}) allowed for the explicit and systematic calculation of Casimir energies and all of the heat-kernel coefficients, which was a long-standing, very hard problem.

We recall that in  gauge theory the small fluctuation operator  $L$ is singular due to  gauge invariance, and therefore a gauge fixing term and the ensuing ghost contributions will necessarily appear. 

The one-loop quantum partition function $Z[L]$, $S_0$ being the classical action, 
\beq
Z[L]\simeq e^{-S_0}\int d[\eta]\,
e^{ -\frac{1}{2} \int d^4x \eta L  \eta},  
\eeq
reduces to a Gaussian functional integral and, as is well known, it can be  
computed in terms of the real eigenvalues, $\lambda_n$, of the operator, namely $L\phi_n=\lambda_n\phi_n$. 
Since $\phi=\sum_n c_n \phi_n $, the formal 
functional measure $d[\phi]$  reads  ($\mu$  is an arbitrary renormalization parameter)
\beq
d[\phi]=\prod_n\frac{dc_n}{\sqrt{\mu}}\,.
\eeq
As a consequence, the one-loop quantum `prefactor' is
\beq
Z_1[L] = \prod_n\frac{1}{\sqrt{\mu}}
\int_{-\infty}^{\infty}dc_n
e^{-\frac{1}{2}\lambda_n c_n^2} =\left[\det(\mu^{-2}{  L})\right]^{-1/2}
\eeq
and the  one-loop Euclidean effective action reads   
\beq
\Gamma_E  =: - \log  Z = S_0+\frac{1}{2}\log ( \det \mu^{-2} {  L})\,. 
\eeq
 The above functional determinant is ill-defined but it can be expressed in the formal way
\beq
\left( \log \det L\right)=-\left(\int_0^\infty dt\,t^{-1} {\mbox{Tr}\,e^{-tL}}\right)\,.
\eeq
For large $t$ one faces no problem, since  $L $  is non negative;  for small $t$ the heat kernel expansion
in the regular smooth case and for $D=4$ reads (see, for example, \cite{G}, and  ref. \cite{byts96})
\beq
\mbox{Tr}\,e^{-tL}\simeq \sum_{r=0}^\infty  A_r t^{r-2}\,.
\eeq
It follows that the formal functional determinant is divergent at $ t=0$, and one  needs  an ulterior regularization.
A simple and useful way to proceed is the use of the dimensional one \cite{dowk}, which in our formulation amounts to the replacement 
\beq
t^{-1} \rightarrow \frac{t^{\varepsilon-1}}{\Gamma(1+\varepsilon)}\,.
\eeq
The related regularized functional determinant, with $\varepsilon$ sufficiently large, is thus
\beq
\log \det L(\varepsilon) = -\int_0^\infty dt\,\frac{t^{\varepsilon-1}}{\Gamma(1+\varepsilon)} \mbox{Tr}\,e^{-tL}=  
-\frac{\zeta(\varepsilon,L)}{\varepsilon}\,,
\eeq
where the generalized zeta function associated with $L$, defined for $Re s > 2$ by
\beq
\zeta(s,L)=\frac{1}{\Gamma(s)}\int_0^\infty dt\,t^{s-1} \mbox{Tr}\,e^{-tL}\,,
\eeq
has been introduced. In order to  be able to handle the cutoff one makes use of the celebrated  theorem by Seeley \cite{seeley}:
``If $L$ is an elliptic differential operator, defined on a smooth and compact manifold, the analytic 
continuation of $\zeta(s,L)$ to the whole complex plane $s$ is regular  at $s=0$.'' 
This provides the zeta-function  determinant \cite{ray,hawk,eli0,klaus}, in the form
\beq
\log \det L = -\zeta'(0,L)\,.
\label{logDet}\eeq
Making use of dimensional regularization, one then arrives at
\beq
\log \det L(\varepsilon) = -\frac{1}{\varepsilon}\zeta(0,L) -\zeta'(0,L)+O(\varepsilon).
\eeq
The computable Seeley-de Witt coefficient $A_2=\zeta(0,L)$ controls the ultraviolet divergence,
while $\zeta'(0,A)$ gives a finite contribution which is, in general, difficult to evaluate, since it is non-local. For examples of exact evaluation see, for instance, \cite{byts96} and references therein.
 
The main aim of the paper is to compute in a closed form the functional determinants related to massive Laplacians on spheres by making use of the non-multiplicative property of zeta-function regularized functional determinants. There are several physical motivations for considering massive Laplacians on spheres.

First, we recall the recent approach presented in Ref. \cite{denef,k}, where the Euclidean partition function associated with a de Sitter black hole is computed making use of the related quasinormal modes.
In this paper,  concrete examples of the very interesting claim \cite{hart0} that black hole quasinormal modes determine the one-loop determinants  have been discussed.

A second example is related to the computation of the one-loop effective action associated with  modified gravitational models on the four dimensional sphere (Euclidean version of the de Sitter space-time) with applications to inflation and dark energy issue \cite{Cogno05,Bamba}. 

Furthermore, other physical applications are mentioned in the recent paper by Dowker, \cite{dowk1}, where a multiplicative anomaly interpretation is also advocated (see also previous references quoted there).  A direct computation appeared also in \cite{sprea1}.

The paper is organized as follows. In section. 2 we will recall some details of the multiplicative anomaly issue. 
In sections 3 and  4,  the factorization technique will be applied to two well known examples: the one-dimensional harmonic oscillator and  QFT at finite temperature. In section. 5 the general case of massive Laplacians on spheres will be studied and section 6 will be devoted to conclusions.  Finally,  appendix A contains explicit expressions for the Laplace operator on spheres in $N=1,2,3,4$ dimensions, and in appendix B the expressions for `vector' and `tensor' Laplacians on the unitary sphere $S^4$ are explicitly computed.

\section{Multiplicative anomaly}

The multiplicative anomaly issue arises most naturally in our approach.
Quite often one has to deal with  products of operators, mainly for convenience, in order to drastically simplify calculations and then the  crucial point arises, that  zeta-function regularized determinants do {\it not} satisfy the multiplicative property, in other words: 
\beq \ln \det (AB)\neq \ln \det A+\ln \det B\,.
\eeq
In fact, generically,  there exists the so-called multiplicative anomaly contribution, defined as the difference
\beq
a(A,B)=\ln \det (AB)-\ln \det (A)-\ln \det (B)\,.  
\eeq
Here it is left understood that the determinants of the two  operators, $A$ and $B$ (which do not exist in a strict sense, being both divergent)
are defined by means of the corresponding zeta-functions. This anomaly was discovered by several authors who had detected the problem independently and came up with particular solutions for each case (sometimes with erroneous results). Wodzicki was the first to give its name to the multiplicative anomaly and to construct a final theory for the whole class of pseudo-differential operators ($\Psi$DO, the ones that appear in all physical applications) by discovering the only trace (up to trivial multiplication) which exists for the whole class.\footnote{And which extends, in a unique way, the celebrated Dixmier trace and the Adler-Manin residue (for a more detailed description see, for instance, \cite{eli1} and references therein).} This trace is now called the {\it Wodzicki residue} and the anomaly can be expressed in terms of it by a very simple general formula, as we will now see.

The multiplicative anomaly can be evaluated by the Wodzicki formula (a discussion can be found in  \cite{guidobook} and references therein).  In the simple but important case in which $A$ and $B$ are two operators
of the same order $a=b$ such a formula becomes
\beq
a(A,B)
=\frac{1}{4b} \, \mbox{res}\left[ (\ln(A B^{-1}))^2 \right]\,,
\label{anomalia}\eeq
where the non-commutative residue $ \mbox{res}\: Q$ related to a 
classical pseudo-differential operator $Q$ of order zero is defined as the coefficient of the logarithmic term in $t$ of the following expansion
\beq
\mbox{Tr}(Qe^{-t H})= \sum_j c_j t^{(j-N)/2}-
\frac{\mbox{res}\:Q}{2} \ln t+O(t \ln t)\,,
\label{res}\eeq
 $H$ being an elliptic non-negative operator of second order (usually the Laplacian is taken), which precise form is irrelevant for the evaluation of $ \res\, Q$.

As already anticipated in this paper we will mainly consider the shift of the Laplace operator (a massive Laplacian) defined on arbitrary dimensional spheres,
and in order to compute the related regularized functional determinants, we shall make use of a product factorization in terms of two first-order operators, $L=AB$, and compute the related functional determinant by the rule
\beq
\ln \det (L)=\ln \det (A)+\ln \det (B)+a(A,B)\,.  
\eeq
In all the cases here considered, the evaluation of functional determinants for the first order operators $A$ and $B$ 
is easier than the evaluation of function determinant of the second-order Laplace-like operator $L$.
Of course there is a price to pay which consists in the computation of 
the multiplicative anomaly $a(A,B)$. This can be evaluated in most physical applications and that is why the anomaly is so important and useful.

\section{Application:  path integral for harmonic oscillator}

As a first example of the advantages of factorization technique, let us consider the one-dimensional harmonic oscillator.
Setting  $\hbar =1$, we can formally write the related Euclidean propagator as 
\beq
 K_T :=\int d[ q]\, e^{- I_E [q]},
\label{propagator}
\eeq
with the Euclidean action given by
\beq
I_E [q] = \int_0^T d t \left(\frac{1}{2} \dot q^2(t) + \frac{\omega^2 q^2(t)}{2} \right)\, \label{euclidean action}.
\eeq
 Here, $[d q]$ represents the formal functional measure, the boundary conditions necessary to give a meaning to a formal path integral. \\
As well known,  the propagator (\ref{propagator}) can be re-written in the  form
\beq
K_T (\mathcal{A}) = \langle q, T \,|\, q_0, =0   \rangle. \label{propagator2}
\eeq
As usual, one may formally proceeds  by splitting $q$ into a `classical' part, 
$q_{cl}$, and a quantum fluctuation $\hat q$, i.e. 
\beq
q(t) := q_{cl}(t) + \hat q(t) \label{splitting}\,.
\eeq  
Here $q_{cl}$ solves the classical equations of motion  obtained by $\delta I_E =0$ with boundary conditions $q(0)=0\,,\, q(T)=q $. Thus, from (\ref{splitting}), it turns out that also the quantum fluctuations have to satisfy 
the boundary conditions $\hat q(0) =0=\hat q(T)$.

The Euclidean action (\ref{euclidean action}) becomes: 
\beq
I_E[q]=I_E[q_{cl}] + \frac{1}{2}\int_0^T d t \;\hat q(t) \left[ - \frac{d^2}{d t^2} +\omega^2\right] \hat q(t) \, \label{step}
\eeq
while the classical action reads
\beq
 I_E[q_{cl}] = \int_0^T d t \left(\frac{1}{2} \dot q_{cl}^2 + \frac{\omega^2}{2} q_{cl}^2  \right) \;,
 \eeq
and it can be easily evaluated.

The propagator assumes the form of a Gaussian integral in the quantum fluctuation variables:
\beq
\langle q ,T \,|\, q_0,0   \rangle = \exp -I_E[q_{cl}] \int [d \hat q] \,\exp -\frac{1}{2}\int_0^T d t \;\hat q(t) 
\left[ - \frac{d^2}{d t^2} +\omega^2\right] \hat q(t). \label{gaussian}
\eeq
As a consequence, one has to give a meaning to the formal Gaussian path integral. To this aim, let us denote by 
\beq
L :=  - \frac{d^2}{d t^2} + \omega^2=L_0+\omega^2 \label{K}
\eeq
the second-order differential operator in $L_2(0,T)$, defined in the dense domain \\ 
$D(L) := \{f, L f \in L_2 (0,T) \;| \}$,  with Dirichlect boundary conditions
$f(0)=f(T)=0$.
 The eigenfunctions are $\sin\left(\pi t/T\right)$ and the the spectrum reads 
\beq
\sigma(L):= l_n := \left(\frac{\pi n}{T}\right)^2 + \omega^2 , \, n=1,2,3,..\;.\label{unphys spectrum}
\eeq

We conclude this Section with the final form of the propagator of the harmonic oscillator obtained
by performing the Gaussian integral (\ref{gaussian}), that is
\beq
\langle q, T \,|\, q_0 , 0   \rangle = N\sqrt{\frac{1}{\det \,L}} \,\exp \left(-I_E[q_{cl}]\right) \;, \label{evaluation}
\eeq   
where $N$ is a normalization constant which can be fixed by requiring that \\
$\langle q ,T=0 \,|\, q_0,0 \rangle=\delta(q-q_0)  $.

\subsection{Evaluation of functional determinant of massive Laplace operator}

The functional determinant of the one-dimensional Laplace operator $L$ may be evaluated by several different techniques. A general one consists in making use of binomial expansion, and expressing the final result as an infinite series of Riemann zeta functions. Another  approach is the use of  the Gelfand-Yaglom-Levit-Smilanski-Forman theorem \cite{GY,LS,F} 
that gives the ratio of two functional determinants associated with ordinary differential operators
 $L$ and $L_0$ in the form
\beq
\frac{\det\,L}{\det \,L_0}=\frac{Y(T)}{Y_0(T)}\,,
\eeq
where $Y(T)$ and $Y_0(T)$ are the solution of the (Cauchy) problem
\beq
LY=0\,, \quad Y(0)=0, \quad \dot Y(0)=1\,, \quad  Y(T)=\frac{\sinh \omega T}{\omega}
\eeq
and
\beq
L_0Y_0=0 \,, \quad Y_0(0)=0, \quad \dot Y_0(0)=1\,, \quad   Y_0(T)=T\,.
\eeq
This leads to the well known result
\beq
\frac{\det \,L}{\det \,L_0}=\frac{\sinh \omega T}{\omega  T}\,.
\label{gel}
\eeq
We will now show that this regularized determinant can be most conveniently computed with 
the factorization technique illustrated in previous section.

To this aim let us first factorize the operator $L$ as 
\beq
L=K^\dag K=(P+i\omega)(P-i\omega)\,,\quad\quad K=P-i\omega\,,
\eeq
where the self-adjoint operator $P=\sqrt{- \frac{d^2}{d t^2}}$ is defined via the spectral theorem of $L_0$ and its spectrum reads 
$\sigma(P)=\frac{\pi n}{T}$, $n=1,2,3,..$.
As a consequence
\beq
\det \, L=\det \,  (K^\dag K)\,,
\eeq
and 
\beq
\ln \det \, L=\ln \det \, K^\dag +\ln \det \, K +a(K^\dag,K)\,.
\eeq
Making us of equations (\ref{anomalia}) and (\ref{res}), a direct calculation leads to $a(K^\dag, K)=0$  
(see also Section 5), 
and thus we have
\beq
\ln \det \, L=-\zeta'(0|K^\dag)-\zeta'(0|K)\,.
\label{sa}
\eeq
Since $\sigma(K^\dag)= \frac{\pi n}{T}+i\omega $, $\sigma(K)= \frac{\pi n}{T}-i\omega $
($n=1,2,3,..$), one easily has
\beq
\zeta(s|K^\dag)(\omega)=\left(\frac{\pi}{T}\right)^{-s}\left[\zeta(s, \frac{i\omega T}{\pi})-\left(\frac{i\omega T}{\pi}\right)^{-s}  \right]\,,
\quad \zeta(s|K)(\omega)=\zeta(s|K^\dag)(-\omega)\,,
\label{s}
\eeq
where $\zeta(s,q)$ is the Hurwitz zeta function defined by  
\beq
\zeta(s,q)=\sum_{n=0}^\infty (n+q)^{-s}\,,\quad\quad\Re s>1\,,\quad q\neq-n\,.
\eeq
The latter expression can be analytical extended to the whole complex $s-$plane  and $q-$plane.
The following properties are valid:
\beq
\zeta(0,q)=\frac{1}{2}-q\,, \quad\quad \zeta'(0,q)=\ln \Gamma(q)-\frac{1}{2}\ln 2\pi\,. 
\label{d}
\eeq
Making use of (\ref{sa}),  (\ref{s}) and (\ref{d}), a direct calculation yields
\beq
 \det \, L= \frac{2\sinh \omega T}{\omega}\,.
\eeq
In the limit $\omega\to0$ goes one obtains
\beq
 \det \, L_0= 2 T\,,
\eeq 
in complete agreement with the G-Y-L-S-F theorem result (\ref{gel}). 

\section{Application: QFT at finite temperature}

As a second example let us consider a free  charged boson field at finite temperature $\beta=1/T$, 
and chemical potential $\mu$.
The related  grand canonical partition function reads
\beq
Z_{\beta,\mu}=\int_{\phi(\tau)=\phi(\tau+\beta)}D\phi_i
e^{-\frac{1}{2}\int_0^\beta d\tau \int d^3x \phi_iA_{ij}\phi_j}
\eeq
with $A_{ij}=\left( L_\tau+ L_3-\mu^2\right)\delta_{ij}+ 2\mu \epsilon_{ij}\sqrt{L_\tau}$, $L_3=-\Delta_3+m^2$,
$\Delta_3$ being the Laplace operator on $R^3$ (it has a continuous
spectrum $\vec k^2$) and
$L_\tau=-\partial^2_\tau$ (it has a discrete spectrum, with Matsubara frequencies
$\omega_n^2=4\pi^2/\beta^2$).
Thus,  the  grand canonical partition function can be written as 
(see, for example, \cite{f} and references therein)
\beq
\ln Z_{\beta,\mu}=-\ln\det \left\| A_{ik}\right\|\,.
\eeq
Now the algebraic determinant of $A$, $|A|$, can be evaluated through the factorization
\beq
| A_{ik}| = \det ( K_+K_-)\,,
\eeq
where $K_\pm=L_3+( \sqrt L_\tau \pm i\mu)^2 $.
However, it is easy to see that another convenient factorization exists \cite{f}, namely 
\beq
| A_{ik}| = \det  ( L_+L_-)\,,
\eeq
with $L_\pm=L_\tau+( \sqrt L_3 \pm \mu)^2$ (again,  $ A = L_+L_-=K_+K_-$), and in both 
cases one is dealing with  the product of two $\Psi$DOs, the  couple $L_+$ and $L_-$ being also
formally self-adjoint. This is a very interesting situation. To wit, the partition function can be written 
in both the forms
\beq
\ln Z_{\beta,\mu} &=&-\ln\det  K_+ -\ln\det K_-+a(K_+,K_-)\,,
\\
&=&-\ln\det  L_+ -\ln\det L_-+a(L_+,L_-)\,.
\eeq
The evaluation of the multiplicative anomalies which appear in both 
expressions above can be performed by making use of  
Wodzicki's formula and, indeed, complete agreement is found for the two expressions
 of the partition function.
Moreover, it is quite easy to realize that if  one neglects the multiplicative anomaly contribution, one arrives at a sound mathematical inconsistency: the results obtained in the two different factorizations are quite different \cite{f}.\footnote{We should again observe that this has caused a substantial number of errors in the literature.}

\section{Functional determinants of massive Laplacians on spheres of arbitrary dimension}

The  multiplicative anomaly plays a relevant role is the case of the evaluation of  functional 
determinants of massive Laplacians on arbitrary dimensional spheres. 
Its specific importance there has also been recently pointed out in references \cite{denef,k} and by Dowker \cite{dea1,dowk1}.   
Our approach in the present paper is however quite different from the ones advocated in \cite{denef,k,dowk1}, in
the sense that we will compute the massive determinant by suitable factorization, 
similar to the one used in the previous section, and compute the associated multiplicative anomaly by making use of Wodzicki's formula.

To start, we recall the eigenvalues and relative degeneration of the Laplace operator acting on scalar function in an $N-$dimensional sphere $S^N$, and we introduce some useful notation, as follows
\beq
\begin{array}{ll}
\lap_N\,,& \mbox{ scalar  Laplacian on }S^N;\\ \\
\la^N_n=n(n+2\nu_N)=(\al^N_n)^2-\nu_N^2\,, & \mbox{ eigenvalues;}\\ \\
d^N_n=\sum_{k=0}^{N-1}\,c^N_k\,(\al^N_n)^k\,,& \mbox{ degeneration;}\\ \\
\Om_N=\frac{2\pi^{(N+1)/2}}{\Ga((N+1)/2)}\,,&        \mbox{volume (hyper-surface) of }S^N; \\ \\
\nu_N=\frac{N-1}2\,,&\al^N_n=n+\nu_N\,,
\end{array}
\label{eigen}\eeq
where $n$ runs from $0$ to $\ii$ and  $(\al^N_n)^2$ are the eigenvalues of the operator $\hat L_N=-\lap_N+\nu_N^2$. 
This is a positive operator and its square-root has eigenvalues  $\al^N_n$ with degeneration $d^N_n$. 

The degeneration of the eigenvalues assumes the explicit form:
\beq
d^N_n=\ag\begin{array}{ll}
d^1_0=1\,,\hs d^1_n=2\,,&d^2_n=2(n+\nu_2)\,,\\ \\
d^N_n=\frac{2}{(N-1)!}\,\prod_{k=0}^{(N-3)/2}\,[(n+\nu_N)^2-k^2]\,,&\mbox{for odd }n\geq3\,,\\ \\
d^N_n=\frac{2(N+\nu_N)}{(N-1)!}\,\prod_{k=0}^{(N-4)/2}\,[(n+\nu_N)^2-(k+1/2)^2]\,,&\mbox{for even }n\geq4\,,
\end{array}\cp\eeq
and this permits to compute the coefficients $c_k^N$ in eq. (\ref{eigen}).

From now on, for simplicity, the index $N$  will be left understood 
(recall all quantities depend on $N$). Thus, we will write 
$L,\nu,c_k,...$ in place of   $L_N,\nu_N,c_k^N,...$, etc. 

We are now ready to compute the determinant of the  Laplace-like operator. For the special operator
$\hat L$, with eigenvalues $\al_n$, we can also compute the corresponding zeta-function in terms of a finite sum of  Hurwitz
zeta-function, but this will not be the case for a Laplacian with arbitrary mass.

For $\Re s$ sufficiently large (this depends  on the dimension), we have
\beq
\ze(s|\hat L)=\sum_{n=0}^\ii\,\sum_{k=0}^{N-1}\,c_k\al_n^{-2s+k}
=\sum_{k=0}^{N-1}\,c_k(n+\nu)^{-2s+k}=\sum_{k=0}^{N-1}\,c_k\ze(2s-k,\nu)\,,
\eeq
where $\ze(s,q)$ is the  Hurwitz zeta-function. 
Note that, for even $N$, $c_{2k}=0$, while for odd $N$, $c_{2k+1}=0$  
and so the sum over $k$ is performed on odd or even $k<N$, only.

Now, let us consider the operator  $L=\hat L+\al^2$ ($\al$ is an  arbitrary constant), 
and the two pseudo-differential operators  $D_\pm$, such that
\beq
L=D_+D_-\,,\hs\hs D_\pm=\sqrt{\hat L}\pm i\al\,.
\eeq
One has
\beq
\log\det L=\log\det D_++\log\det D_-+a(D_+,D_-)\,,
\eeq
where $a(D_+,D_-)$ is the multiplicative anomaly. In order to compute it, we make use of  Eqs. (\ref{anomalia}) and (\ref{res}),  
choosing $H=\hat L$ in (\ref{res}), the spectral theorem gives
\beq
\mbox{Tr}\at\aq\log\frac{\sqrt{\hat L}+i\al}{\sqrt{\hat L}-i\al}\cq^2\,e^{-t\hat L}\ct
&=&\sum_{n=0}^\ii\,\e^{-t\al_n^2}f(\al_n)\,
\nn\\
&=&\sum_j c_j t^{(j-N)/2}-\frac{\mbox{res}\:\aq\log\frac{\sqrt{\hat L}+i\al}{\sqrt{\hat L}-i\al}\cq^2\    }{2} \ln t+O(t \ln t)\,,
\eeq
where 
\beq
f(\al_n)=d_n^N(\al_n)\,\at\log\frac{\al_n+i\,\al}{\al_n-i\al}\ct^2\,.
\eeq
In this way 
\beq
\res\,Q=-\Res(f(z),z=\ii)\,,
\eeq
$\Res(f,z)$ being the ordinary Cauchy residue. As a result, the multiplicative anomaly  reads     
\beq
a(D_+,D_-)=-\frac{\Res(f(z),z=\ii)}4\,.
\label{anCau}\eeq
In odd dimensions, $d_n^N$ is an even polynomial in $\al_n$ and so the multiplicative anomaly 
is trivially vanishing, while in even dimensions the multiplicative anomaly 
is a polynomial of order $N$.

Using  (\ref{logDet}) for the regularized definition of the determinant of the operator we get, in the present case,
\beq
\log\det L=\log\det(D_+D_-)=-\ze'(0|D_+)-\ze'(0|D_-)+a(D_+,D_-)\,.
\eeq
For $\Re s$  sufficiently large (the actual value depends on the dimension), we obtain
 \beq
\ze(s|D_\pm)&=&\sum_{n=0}^\ii\,\sum_{k=0}^{N-1}\,c_k\,\al_n^k(\al_n\pm i\al)^{-s}
\nn\\
&=&\sum_{n=0}^\ii\,\sum_{k=0}^{N-1}\,\sum_{j=0}^k\,c_k\,b_{kj}
(\al_n\pm i\al)^j(\mp i\al)^{k-j}(\al_n\pm i\al)^{-s}
\nn\\
&=&\sum_{k=0}^{N-1}\sum_{j=0}^k\,c_k\,b_{kj}(\mp i\al)^{k-j}\,\ze(s-j,\nu\pm i\al)\,,
\eeq
where $b_{kj}$ are binomial coefficients. Then, it finally follows that
\beq
\log\det L&=&a(D_+,D_-)
              \nn\\&& -\sum_{k=0}^{N-1}\sum_{j=0}^k\,c_kb_{kj}(i\al)^{k-j}\,
                                       \aq\ze'(-j,\nu-i\al)+(-1)^{k-j}\,\ze'(-j,\nu+i\al)\cq\,.
\eeq
This general expression yields the logarithm of the determinant of the massive Laplace operator on 
an hypersphere in any number of dimensions in terms of  a {\it finite} sum of Hurwitz zeta-functions. 
In App. \ref{app1} we  write explicit formulae for dimensions $N=1,2,3,4$,
while in App. \ref{app2} we extend the computation to Laplace-Beltrami operators acting 
on vector and tensor fields. In these last cases, we limit ourselves to the physical dimension $N=4$.

Finally, with regard to the comparison with other works, our results agree with 
similar results obtained by Dowker in \cite{dea1,dowk1}.  
 
\section{Conclusions}

At the beginning of this paper we have presented a brief survey of zeta function regularization and, in especial, of the related multiplicative anomaly issue, what has been done in Sect. 2. Our point being that, as can be easily checked on the physical literature on the subject, even if those issues are since long well and rigorously established in the mathematical community dealing with physical applications, this is still not the case among physicists. It is worthwhile to summarize the main points again and explain them with the help of a couple of basic but very useful examples, namely the one-dimensional harmonic oscillator and  QFT at finite temperature, as we did in Sects. 3 and 4, respectively.

A major new contribution has been the computation, carried out in Sect. 5, of the general case corresponding to massive Laplacians on spheres in arbitrary dimensions.
A general formula which yields the determinant of the massive Laplace operator on 
a hypersphere in {\it any} number of dimensions in terms of  a {\it finite} sum of Hurwitz zeta-functions has been obtained. Further, explicit expressions for the Laplace operator on spheres in $N=1,2,3,4$ dimensions have been given, in App. A, and those for `vector' and `tensor' Laplacians on the unitary sphere $S^4$, in App. B.
These last are, in no way, straightforward cases, and the fact that could 
be treated in such simple and general way by using the zeta function procedure 
is an excellent proof of the power of this method. 

The results concerning the scalar Laplacian
agree with analogue results  obtained in \cite{dea1,dowk1}.

\medskip

\noindent {\bf Acknowledgements.} 

We would like to thank Prof. Stuart Dowker for pointing out to us an error in the preliminary version of the paper.

E.E. was supported in part by MINECO (Spain), project FIS2010-15640, and by the CPAN Consolider Ingenio Project.

\appendix
\s{Laplace operator on sphere in $N=1,2,3,4$ dimensions}
\label{app1}
Consider the operators $L=-\lap+\rho^2$ acting on functions on the sphere in 1,2,3 and 4 dimensions.
Here we use the notation in Section 5.

\ss{$N=1$}
For eigenvalues and degeneration one  trivially has  
$\la_n=n^2$, $d_n=2$ (excluding the vanishing eigenvalue $\la_0=0$). 
The multiplicative anomaly is vanishing and so
\beq
\ze(s|\hat L_1)=2\ze(2s)\,,\hs\hs\al=\rho\,,
\eeq
\beq
\log\det L_1&=&-2\aq\ze'(0,-i\al)+\ze'(0,i\al)\cq\,.
\eeq 

\ss{$N=2$}
\beq
L_2=\hat L_2+\al^2\,,\hs\hs \al^2=\rho^2-\nu^2_2=\rho^2-\frac14\,.
\eeq
\beq
\nu=\frac12\,,\hs d_n=2(n+\nu)\,,\hs c_1=2\,,\hs c_k=0\mbox{ for }k\neq1\,,
\eeq
\beq
\ze(s|\hat L_2)=2\ze(2s-1,1/2)=2\at2^{2s-1}-1\ct\,\ze(2s-1)\,,
\eeq 
\beq
\log\det L_2&=&-2\al^2-2i\al\,\aq\ze'(0,1/2-i\al)-\ze'(0,1/2+i\al)\cq
\nn\\&&\hs\hs-2\aq\ze'(-1,1/2-i\al)+\ze'(-1,1/2+i\al)\cq\,.
\label{N=2}\eeq 

\ss{$N=3$}
\beq
L_3=\hat L_3+\al^2\,,\hs\hs \al^2=\rho^2-\nu^2_3=\rho^2-1\,,
\eeq
\beq
\nu=1\,,\hs d_3=2(n+1)\,,\hs c_2=1\,,\hs c_k=0\mbox{ for }k\neq2\,,
\eeq
\beq
\ze(s|\hat L_3)=\ze(2s-2,1)=\ze(2s-2)\,,
\eeq 
\beq
\log\det L_3&=&\al^2\,\aq\ze'(0,1-i\al)+\ze'(0,1+i\al)\cq
\nn\\&&\hs\hs-2i\al\,\aq\ze'(-1,1-i\al)-\ze'(-1,1+i\al)\cq
\nn\\&&\hs\hs\hs-\aq\ze'(-2,1-i\al)+\ze'(-2,1+i\al)\cq
\label{N=3}\eeq 

\ss{$N=4$}
\label{N=4}
Here $\al^2=\rho^2-9/4$ and
\beq
\ze(s|\hat L_4)=-\frac1{12}\,\ze(2s-1,3/2)+\frac13\,\ze(2s-3,3/2)\,,
\eeq
\begin{eqnarray}
\log\det L_4&=&\frac29\,\al^4+\frac1{12}\,\al^2
+\frac{1}{3}i\al^3\,\aq\zeta'\left(0,3/2-i\al\right)-\zeta'\left(0,3/2+i\al\right)\cq\nonumber\\&&\hs\hs
+\frac{1}{12}i\al\,\aq\zeta'\left(0,3/2-i\al\right)-\zeta'\left(0,3/2+i\al\right)\cq\nonumber\\&&\hs\hs\hs
+\al^2\,\aq\zeta'\left(-1,3/2-i\al\right)+\zeta'\left(-1,3/2+i\al\right)\cq\nonumber\\&&\hs\hs\hs\hs\
+\frac{1}{12}\,\aq\zeta'\left(-1,3/2-i\al\right)+\zeta'\left(-1,3/2+i\al\right)\cq\nonumber\\&&\hs\hs\hs
-i\al\,\aq\zeta'\left(-2,3/2-i\al\right)-\zeta'\left(-2,3/2-i\al\right)\cq\nonumber\\&&\hs\hs
-\frac{1}{3}\,\aq\zeta'\left(-3,3/2-i\al\right)+\zeta'\left(-3,3/2+i\al\right)\cq\,.
\end{eqnarray}

\section{`Vector' and `Tensor' Laplacian on unitary sphere $S^4$}
\label{app2}

Now we consider the Laplace-Beltrami operators $\lap^{(v,t)}$ acting on traceless-transverse vector
and tensor fields on the unitary sphere $S^4$. In such a case the eigenvalues and the
corresponding degenerations read  
\beq
\begin{array}{ll}
\lap^{(v,t)}\,,& \mbox{ vector/tensor  Laplacian on }S^4;\\ \\
\la^{(v,t)}_n=[n+\nu_{(v,t)}]^2-\ga_{(v,t)}\,, & \mbox{ eigenvalues;}\\ \\
d^{(v,t)}_n=a^{(v,t)}[n+\nu_{(v,t)}]+b^{(v,t)}[n+\nu_{(v,t)}]^3 \,,& \mbox{ degeneration;}\\ \\
\nu_{(v)}=\frac52\,,&\nu_{(t)}=\frac72\,,\\ \\
\ga_{(v)}=\frac{13}4\,,&\ga_{(t)}=\frac{17}4\,,\\ \\
a^{(v)}=-\frac94\,,& b^{(t)}=1\,,\\ \\
a^{(t)}=-\frac{125}{12}\,,& b^{(t)}=\frac53\,,
\end{array}
\eeq
$[n+\nu_{(v,t)}]^2$ being the eigenvalues of the operators 
$\hat L^{(v,t)}=-\lap^{(v,t)}+\ga_{(v,t)}$. 
These are positive operators and their square-roots 
have eigenvalues  $n+\nu_{(v,t)}$ with degenerations $d^{(v,t)}_n$. 

Now we can proceed as in Section 5  and define
\beq
L^{(v,t)}&=&-\lap^{(v,t)}+\rho^2=L^{(v,t)}+\al_{(v,t)}^2=D^{(v,t)}_+D^{(v,t)}_-\,,
\eeq
\beq
\log\det L^{(v,t)}&=&-\ze'(0|D^{(v,t)}_+)-\ze'(0|D^{(v,t)}_-)+a(D^{(v,t)}_+,D^{(v,t)}_-)
\nn\\
&=&-\ze'(0|D^{(v,t)}_+)-\ze'(0|D^{(v,t)}_-)+a(D^{(v,t)}_+,D^{(v,t)}_-)\,,
\eeq
where 
\beq
\al_{(v)}^2=\rho^2-\ga_{(v)}=\rho^2-\frac{13}4\,,\hs\hs
 \al_{(t)}^2=\rho^2-\ga_{(t)}=\rho^2-\frac{17}4\,.
\eeq
Also in these cases the anomaly can be easily computed by means of equation (\ref{anCau}, with 
appropirate eigenvalues and degeneration.  

For simplicity here we assume positive eigenvalues of the operators $L^{(v,t)}$. 
Negative or vanishing eigenvalues have to be considered separately 
(see appendix in \cite{Cogno05}).

The specific computation is similar to the one in example \ref{N=4}. One gets 
\begin{eqnarray}
\zeta(0|L^{(v)})&=&\frac14\,\rho^4-\frac12\,\rho^2-\frac{15}4\,,
\end{eqnarray}
\begin{eqnarray}
\log\det L^{(v)}&=&\frac23\,\alpha^4+\frac94\,\al^2
   +i{\ze}'\left(0,\frac{5}{2}-i\alpha\right)\alpha^3
    -i{\ze}'\left(0,\frac52+i\al\right)\alpha^3
\nn\\&&\hs
   +3{\ze}'\left(-1,\frac{5}{2}-i\alpha\right)\alpha^2
     +3{\ze}'\left(-1,\frac52+i\al\right)\alpha^2
\nn\\&&\hs\hs
   -3i{\ze}'\left(-2,\frac{5}{2}-i\alpha\right)\alpha
    +3 i{\ze}'\left(-2,\frac52+i\al\right)\alpha
\nn\\&&\hs\hs\hs
    +\frac{9}{4}i{\ze}'\left(0,\frac{5}{2}-i\alpha\right)\alpha
      -\frac{9}{4}i{\ze}'\left(0,\frac52+i\al\right)\alpha
\nn\\&&\hs\hs
   -{\ze}'\left(-3,\frac{5}{2}-i\alpha\right)
     -{\ze}'\left(-3,\frac52+i\al\right)
\nn\\&&\hs
 +\frac{9}{4}{\ze}'\left(-1,\frac{5}{2}-i\alpha\right)
   +\frac{9}{4}{\ze}'\left(-1,\frac52+i\al\right)\,,
\label{vec}\end{eqnarray}
\begin{eqnarray}
\zeta(0|L^{(t)})=&=& \frac5{12}\,\rho^4+\frac53\,\rho^2-\frac{40}{3}\,,
\end{eqnarray}
\begin{eqnarray}
\log\det L^{(t)}&=& \frac{10}9\,\alpha^4+\frac{125}{12}\,\al^2
   +\frac{5}{3}i{\ze}'\left(0,\frac{7}{2}-i\alpha\right)\alpha^3
    -\frac{5}{3}i{\ze}'\left(0,\frac72+i\al\right)\alpha^3
\nn\\&&\hs
   +5{\ze}'\left(-1,\frac{7}{2}-i\alpha\right)\alpha^2
    +5{\ze}'\left(-1,\frac72+i\al\right)\alpha^2
\nn\\&&\hs\hs 
  -5 i{\ze}'\left(-2,\frac{7}{2}-i\alpha\right)\alpha
    +5 i{\ze}'\left(-2,\frac72+i\al\right)\alpha
\nn\\&&\hs\hs\hs 
  +\frac{125}{12}i{\ze}'\left(0,\frac{7}{2}-i\alpha\right)\alpha
    -\frac{125}{12}i{\ze}'\left(0,\frac72+i\al\right)\alpha
\nn\\&&\hs\hs
   -\frac{5}{3}{\ze}'\left(-3,\frac{7}{2}-i\alpha\right)
    -\frac{5}{3}{\ze}'\left(-3,\frac72+i\al\right)
\nn\\&&\hs 
   +\frac{125}{12}{\ze}'\left(-1,\frac{7}{2}-i\alpha\right)
     +\frac{125}{12}{\ze}'\left(-1,\frac72+i\al\right)\,.
\label{tens}\end{eqnarray}
It is understood that in all equations above, $\al$ has to be replaced by the 
appropriate expression ($\al_v$ or $\al=\al_t$).


\end{document}